\documentclass[12pt]{article}
\usepackage{epsfig}
\newcommand{\beq}{\begin{equation}}
\newcommand{\eeq}{\end{equation}}
\def\lsim{\:\raisebox{-0.5ex}{$\stackrel{\textstyle<}{\sim}$}\:}
\def\gsim{\:\raisebox{-0.5ex}{$\stackrel{\textstyle>}{\sim}$}\:}

\title{\Large\bf Born-form approximation and full one-loop results for 
\boldmath$e^+ e^- \to W^+W^- \to$\unboldmath 4 fermions 
\boldmath$(+ \gamma)$\unboldmath\footnote{Presented by D. Schildknecht at 
the ECFA/DESY Linear Collider
Workshop, Obernai, 16-19th October, 1999, to appear in the Workshop
Proceedings.} \footnote{Supported by the Bundesministerium f{\"u}r 
Bildung und Forschung, No.05HT9PBA2, Bonn, Germany}}
\author{{\bf Masaaki Kuroda}\\
Institute of Physics, Meiji-Gakuin University,\\
Yokohama 244, Japan
\bigskip\\
and
\bigskip\\
{\bf Dieter Schildknecht}\\
Department of Physics, University of Bielefeld,\\ 
Universit{\"a}tsstra{\ss}e 25, D-33615 Bielefeld, Germany}
\date{}

\begin{document}
\maketitle
\vspace*{4em}
\begin{abstract}
We review the results on representing the differential cross section for 
W-pair production, including W decay and hard-photon bremsstrahlung, in terms
of a Born-form approximation of fairly simple analytic form. The results
of the Born-form approximation are compared with full-one-loop results. 
The emphasis is on the energy range of future $e^+e^-$ (or $\mu^+\mu^-$)
colliders.
\end{abstract}
\vfill
LC-TH-2000-018 \\
BI-TP 2000/05 \\
hep-ph/0002099
\vspace*{2em}

\clearpage
  
\section{Introduction}
I will start with a few remarks on W-pair production at LEP energies based 
on work \cite{Ku-Ku} in collaboration with Masaaki Kuroda and Ingolf Kuss.
Subsequently, I will turn to W-pair production at the energy range of future
$e^+e^-$ (or $\mu^+ \mu^-$) colliders. Specifically, the talk will
be based on recent work in collaboration with Masaaki Kuroda and 
Yoshimasa Kurihara 
\cite{Ku-Schi,Kuri,Polonica}. The high-energy-Born-form
approximation (HEBFA) is compared with full-one-loop results including
W-decay. The double-pole approximation employed throughout is explicitly
justified by demonstrating its validity, provided appropriate cuts are
introduced on the two-fermion invariant masses. As a consequence of the
strong increase with energy of the virtual corrections involving the
non-Abelian couplings, even limited experimental accuracy at TeV energies
will be seen to be sufficient to isolate their effects. For elaborate work
on W-pair production, on various aspects more complete than the work
presented here, I would like to refer to the talks by A. Denner \cite{DDRW}
and A. Vicini at this meeting.

\section{The Born Approximation}
The Born approximation for the reaction $e^+ e^- \to W^+ W^-$, based on
the well-known s-channel, $(\gamma, Z_0)$-exchange and t-channel,
$\nu$-exchange diagrams may be written as (e.g. \cite{Ku-Ku})
\beq
{\cal M}_{Born} (\sigma, \lambda_+, \lambda_-, s,t) = g^2_{W^\pm} {1 \over 2}
\delta_{\kappa^-} {\cal M}_I + e^2 {\cal M}_Q,
\eeq
where the dependence on energy and momentum transfer squared, s and t, and
on twice the electron and the $W^\pm$ helicities, $\sigma = \pm 1$ 
and $\lambda_\pm
= 0, \pm 1$, is contained in the basic matrix elements ${\cal M}_I$ and
${\cal M}_Q$. The calculation of the cross section for $e^+e^- \to
W^+W^-$ in (1) requires the specification of an appropriate energy scale
at which the SU(2) coupling, $g_{W^\pm}$ and the electromagnetic coupling,
$e$ are to be defined. For W-pair production at LEP2 energies of $2M_{W^\pm}
\lsim \sqrt s \lsim 200 GeV$, it is natural to chose a high-energy scale,
such as $\sqrt s$, or, with sufficient accuracy, $M_W \simeq M_Z$  instead
of $\sqrt s$. Accordingly, we have
\beq
\left({{e^2} \over {4 \pi}}\right)^{-1} = \alpha^{-1} (M^2_Z) = 128.89 \pm 0.09
\eeq
for the electromagnetic coupling, while $g^2_{W^\pm} (M^2_W)$ is obtained
\cite{Ku-Ku} from the leptonic width of the $W^\pm$
\beq
g^2_{W^\pm} \left(M^2_{W^\pm} \right) = 48 \pi {{\Gamma^W_e} \over {M_{W^\pm}}}.
\eeq
The $W^\pm$ width not being known experimentally with sufficient accuracy, the
theoretical one-loop expression for the leptonic W width in terms of the
well-measured Fermi coupling, $G_\mu$, from $\mu^\pm$ decay
\beq
\Gamma^W_e = {{G_\mu M^3_W} \over {6 \sqrt 2 \pi (1 + \Delta y^{SC})}}
\eeq
is to be inserted in (3) to yield \cite{Ditt}
\beq
g^2_{W^\pm} \left(M^2_{W^\pm}\right) = {{4 \sqrt 2 G_\mu M^2_{W^\pm}} 
\over {1 + \Delta
y^{SC}}}.
\eeq
The one-loop correction, $\Delta y^{SC}$, where SC stands for the change
of scale between $\mu$-decay and W-decay, amounts to \cite{Ku-Ku}
\begin{eqnarray}
\Delta y^{SC} & = & \Delta y^{SC}_{ferm} + \Delta y^{SC}_{bos}\nonumber \\
& = & (-7.79 + 11.1) \times 10^{-3} = 3.3 \times 10^{-3}.
\end{eqnarray}
The numerical value is practically independent of the Higgs-boson mass. As
indicated in (6), there is a significant cancellation between bosonic and
fermionic corrections operative in $\Delta y^{SC}$.
\begin{table}[tbp]
\footnotesize
\begin{center}
\vspace{-5mm}
\begin{tabular}{|r|r|r|c||r|r|c|}
\hline
\multicolumn{1}{|c|}{angle}&\multicolumn{3}{c||}{unpolarized}&
\multicolumn{3}{c|}{left-handed}\\
\hline
&\multicolumn{1}{c|}{$\Delta_{IBA}$}&$\delta\Delta_{IBA}$&
\multicolumn{1}{c|}{$\Delta_{
IBA}+\delta\Delta_{IBA}$}&
$\Delta_{IBA}$&$\delta\Delta_{IBA}$&
\multicolumn{1}{c|}{$\Delta_{IBA}+\delta\Delta_{IBA}$}\\
\hline\hline
\multicolumn{7}{|c|}{$\sqrt{s}=161$ GeV}\\
\hline
\multicolumn{1}{|c|}{total}&1.45&-0.72&0.73&1.45&-0.72&0.73\\
10&1.63&-0.73&0.90&1.63&-0.73&0.90\\
90&1.44&-0.72&0.72&1.44&-0.72&0.72\\
170&1.26&-0.70&0.56&1.26&-0.70&0.56\\
\hline\hline
\multicolumn{7}{|c|}{$\sqrt{s}=165$ GeV}\\
\hline
\multicolumn{1}{|c|}{total}&1.27&-0.71&0.56&1.28&-0.71&0.57\\
10&1.67&-0.74&0.93&1.67&-0.74&0.93\\
90&1.17&-0.71&0.46&1.18&-0.71&0.47\\
170&0.75&-0.67&0.08&0.77&-0.67&0.10\\
\hline\hline
\multicolumn{7}{|c|}{$\sqrt{s}=175$ GeV}\\
\hline
\multicolumn{1}{|c|}{total}&1.26&-0.71&0.55&1.28&-0.71&0.57\\
10&1.71&-0.75&0.96&1.71&-0.75&0.96\\
90&1.03&-0.69&0.34&1.06&-0.70&0.36\\
170&0.59&-0.62&-0.03&0.69&-0.63&0.06\\
\hline\hline
\multicolumn{7}{|c|}{$\sqrt{s}=184$ GeV}\\
\hline
\multicolumn{1}{|c|}{total}&1.02&-0.70&0.32&1.06&-0.71&0.35\\
10&1.57&-0.75&0.82&1.57&-0.75&0.82\\
90&0.67&-0.68&-0.01&0.72&-0.69&0.03\\
170&0.10&-0.58&-0.48&0.32&-0.64&-0.32\\
\hline\hline
\multicolumn{7}{|c|}{$\sqrt{s}=190$ GeV}\\
\hline
\multicolumn{1}{|c|}{total}&1.24&-0.70&0.54&1.28&-0.71&0.57\\
10&1.67&-0.74&0.93&1.67&-0.75&0.92\\
90&0.95&-0.68&0.27&1.01&-0.69&0.32\\
170&0.58&-0.57&0.01&0.83&-0.59&0.24\\
\hline\hline
\multicolumn{7}{|c|}{$\sqrt{s}=205$ GeV}\\
\hline
\multicolumn{1}{|c|}{total}&1.60&-0.70&0.90&1.65&-0.71&0.94\\
10&1.77&-0.74&1.03&1.77&-0.74&1.03\\
90&1.55&-0.66&0.89&1.64&-0.68&0.96\\
170&1.61&-0.53&1.08&1.94&-0.56&1.38\\
\hline
\end{tabular}
\end{center}

\caption{
  The Table shows the quality of the improved Born approximation (IBA) for the
  total (defined by integrating over $10^0\protect\lsim\vartheta\protect\lsim 
  170^0$) and the
  differential cross section (for $W^-$-production angles $\vartheta$ of
  $10^0,90^0$ and $170^0$) for $e^+e^-\protect\to W^+W^-$ 
  at various energies for
  unpolarized and left-handed electrons. The final percentage deviation,
  $\Delta_{IBA} + \delta \Delta_{IBA}$, of the IBA from the full one-loop 
  result
  is obtained by adding the correction $\delta \Delta_{IBA}$ resulting from
  using the appropriate high energy scale in the SU(2) coupling, to the
  percentage deviation, $\Delta_{IBA}$, based on using the low-energy scale in
  the SU(2) coupling, i.e. $\Delta y^{SC} =0$. (From \cite{Ku-Ku})}

\end{table}

\section{The improved Born approximation at LEP2.}
Supplementing the Born approximation (1), with the coupling constants
from (2) and (5), by a Coulomb correction and by initial-state radiation
(ISR) in soft-photon approximation \cite{Boehm,Been}, the improved
Born approximation for LEP2 energies takes the form \cite{Ku-Ku}
\begin{eqnarray}
\left( {{d \sigma} \over {d \Omega}} \right)_{IBA}
& = & {\beta \over {64 \pi^2s}} \left| {{2 \sqrt 2 G_\mu M^2_W} \over
{1 + \Delta y^{SC}}} {\cal M}^\kappa_I \delta_{\kappa^-} + 4 \pi \alpha
(M^2_Z) {\cal M}^\kappa_Q \right|^2 \nonumber \\
& + & \left({{d \sigma} \over {d \Omega}} \right)_{Coul} (1 - \beta^2)^2
+ \left( {{d \sigma} \over {d \Omega}} \right)_{ISR}.
\end{eqnarray}
A detailed numerical comparison between the full one-loop results and
the results from the simple representation (7) was carried out in ref.
\cite{Been} (without the correction $\Delta y^{SC}$) and in ref. \cite{Ku-Ku}
(taking into account $\Delta y^{SC}$). The results are presented in
Table 1.

The Table shows the percentage deviation of the IBA (7) from the full-one-loop
results for $\Delta y^{SC} = 0$, denoted by $\Delta_{IBA}$, and upon including
$\Delta y^{SC} \not= 0$, denoted by $\Delta_{IBA} + \delta \Delta_{IBA}$.
Upon including the correction due to $\Delta y^{SC}$ from (6), the
deviations of the improved Born approximation from the full one-loop
results are less than 1 \% in the full angular range of the production
angle between 10 degrees and 170 degrees.

We note that the effect of $\Delta y^{SC}$ on the cross section can be 
easily estimated. The cross section (7) being dominated by the part
proportional to ${\cal M}_I$, upon neglecting ${\cal M}_Q$ in (7), one
obtains
\beq
\delta \Delta_{IBA} \simeq - 2 \Delta y^{SC} = - 0.66 \%.
\eeq
This value approximately coincides with the (production-angle-dependent)
results in Table 1.

We finally comment on the significance of the appropriate choice of
the high-energy scale in the weak coupling, $g_W^\pm (M^2_W)$, with 
respect to recent one-loop calculations \cite{Been-al} 
which incorporate the decay
of the $W^\pm$ into 4 fermions in a gauge-invariant formulation.
These calculations take into account fermion-loops only. While
interesting as a first step towards a full one-loop evaluation
of $e^+ e^- \to 4$ fermions, the numerical results of a calculation
including fermion loops only can easily be estimated within the
present framework of stable $W^\pm$ to enlarge the cross section
appreciably. In fact, taking into account
fermion loops only, the estimate (8) changes sign and becomes
\beq
\delta\Delta_{IBA}\vert_{ferm}\simeq -2\Delta y^{SC}_{ferm} 
\vert_{m_t = 180 GeV} \simeq +1.56\%, 
\eeq
and the total deviation from the full one-loop results (using
$\Delta_{IBA} \simeq 1.2 \%$ from Table 1) rises to values of
\beq
\Delta_{IBA} + \delta \Delta_{IBA}\vert_{ferm}\cong 2.8 \% .
\eeq
Accordingly, results from fermion-loop calculations including the
decay of the $W^\pm$ are expected to overestimate the cross section 
by almost 3 \%
relative to the (so far unknown) outcome of a complete calculation of 
$e^+e^- \to 4$ fermions including bosonic loops as well. It is
gratifying, that a simple procedure immediately suggests itself for
improving the large discrepancy (10). One simply has to approximate
the bosonic loop corrections by using the substitution 
\beq
G_\mu\to G_\mu/(1+\Delta y^{SC}_{bos})
\eeq
with $\Delta y^{SC}_{bos} = 11.1 \times 10^{-3}$ in the four-fermion
production amplitudes. Substitution (11) practically amounts to
using $g_{W^\pm} (M^2_W)$ in four-fermion production as well.
With substitution (11), it is indeed to be expected that the 
deviation of four-fermion production in the fermion-loop scheme
will be diminished from the above estimated value of $\simeq 2.8 \%$
to a value below 1 \%.

\section{The high-energy-Born-form approximation (HEBFA) for \boldmath $e^+e^-
\to W^+W^-$\unboldmath at one loop.}

I turn to W-pair production in the high-energy region to be explored
by a future $e^+e^-$ linear collider or by a $\mu^+\mu^-$ collider.
The subsequent HEBFA will turn out to be valid at center-of-mass energies
above a lower limit of approximately 400 GeV.

Including one-loop corrections \cite{Jeger,Sack}, the helicity amplitudes
for $e^+e^-$ annihilation into W-pairs may be represented in terms
of twelve invariant amplitudes
\begin{eqnarray}
{\cal H} (\sigma, \lambda, \bar \lambda) & = & S_I^{(\sigma)} (s,t)
{\cal M}_I (\sigma, \lambda, \bar \lambda) + S_Q^{(\sigma)} (s,t)
{\cal M}_Q (\sigma, \lambda, \bar \lambda) \nonumber \\
& + & \sum_{i = 2,3,4,6} Y_i^{(\sigma)} (s,t) {\cal M}_i (\sigma, \lambda,
\bar \lambda).
\end{eqnarray}
The structure of the electroweak theory, its renormalizability in particular,
restricts (renormalized) ultraviolet and infrared divergences to only affect
the invariant amplitudes, $S_I^{(\sigma)}$ and $S_Q^{(\sigma)}$, multiplying
the basic matrix elements that are also present in the Born approximation.
Accordingly, it is suggestive to approximate \cite{Boehm,Kneur} the 
helicity amplitudes (12) in the high-energy limit by dropping all
contributions in (12) beyond the ones with a structure identical to
the Born approximation. As the bosonic matrix elements do not form an
orthonormal vector space, this requirement does not uniquely determine
$S_I^{(\sigma)} (s,t)$ and $S_Q^{(\sigma)} (s,t)$. Motivated by the 
necessary condition of unitarity constraints at high energies, a certain
choice of the basic matrix elements was suggested and numerically
explored in the early nineties in refs. \cite{Boehm,Kneur}. More recently
it was shown \cite{Ku-Schi} that a somewhat different choice of the basic
matrix elements has the advantage of reproducing the amplitudes for 
production of longitudinal W bosons exactly in the HEBFA. As the production
of longitudinal W bosons is dominant at high energies, this novel choice
is to be the preferred one.

Moreover, while the previous analysis \cite{Boehm,Kneur} of the validity
of the HEBFA was carried out purely numerically, in the more recent paper
\cite{Ku-Schi} simple analytical formulae for the invariant amplitudes
$S^{(-)}_I$ and $S^{(\pm)}_Q$ in (12) were presented. Upon including
soft-photon radiation, the invariant amplitudes $S^{(-)}_I$ and $S^{(\pm)}_Q$ 
turn into \cite{Ku-Schi}
\beq
\hat S_{(I,Q)} = S_{(I,Q)} |_{\Delta \alpha, m_t}
+ S_{(I,Q)} (\Delta E) |_{brems} + S^{dom}_{(I,Q)}.
\eeq
The first term in (13) contains the running of the electromagnetic
coupling and the SU(2) breaking due to the top quark. The second term
is due to the soft-photon bremsstrahlung, while the third one contains
the remaining non-universal loop corrections, in particular all the bosonic
loops, in high-energy approximation.

The analytic expressions for $S^{dom}_{(I,Q)}$ were extracted from
ref. \cite{Denner}, where cross sections for W-pair production for various
helicities were deduced in a systematic high-energy expansion, without
the attempt of constructing a Born-form approximation. Replacing
subdominant terms (that fill several pages of formulae in ref. \cite{Denner})
by constants, the expressions deduced
for $S^{dom}_{I,Q}$ fit on less than two
pages and are presented below:
\footnotesize
\begin{eqnarray}
   S_I^{(-)dom}
     & = & {{\alpha}\over{4\pi s_W^2}}
     \Biggl[-{{1+2c_W^2+8c_W^4}\over{4c_W^2}}(\log{s\over{M_W^2}})^2
       + (4+2{s\over u})(\log{s\over{M_W^2}})(\log{s\over t}) \nonumber \\
     & - & ({{s[s(1-6c_W^2)+3t]}\over{4c_W^2(t^2+u^2)}}+
          {{s(1-6c_W^2)}\over {2c_W^2 u}}) (\log{s\over t})^2 \nonumber \\
     & - &{{3st}\over{2(t^2+u^2)}}(\log{s\over u})^2
         -{{2s}\over u}(\log{s\over t})(\log{s\over u}) \nonumber \\
     & + &{{3(s_W^4+3c_W^4)}\over{4c_W^2}}\log{s\over {M_W^2}}
        - {{1-4c_W^2+8c_W^4}\over{2c_W^2}}(\log{s\over{M_W^2}})(\log c_W^2)
           \nonumber \\
     & + &2(1-2c_W^2)(\log{t\over u})(\log {s\over{M_Z^2}})
          -2s_W^2(\log {t \over u})^2 -8s_W^2Sp(-{u\over t}) \nonumber \\
     & - &{{s[3s+t+6c_W^2(s+3t)]}\over{4c_W^2(t^2+u^2)}}\log{s\over t}
         -{{(1-6c_W^2)su}\over{4c_W^2(t^2+u^2)}} \Biggr] - 0.012, 
\end{eqnarray}

\begin{eqnarray}
     S_Q^{(-)dom} & = & {{\alpha}\over{8\pi s_W^2}}
    \Biggl [-{{3-4c_W^2+12c_W^4-16c_W^6}\over{4c_W^2s_W^2}}(\log{s\over{M_W^2}})^
2
          \nonumber \\
    & + & {{56-97c_W^2+76c_W^4-36c_W^6}\over{6c_W^2s_W^2}}\log{s\over{M_W^2}}
         \nonumber \\
     & - & (1-2c_W^2){{2(1-2c_W^2)^2+1}\over{2c_W^2s_W^2}}\log c_W^2
          \log {s\over{M_W^2}}
        + (4 + 2{{1-2c_W^2}\over{s_W^2}}~{s\over u})\log{s\over{M_W^2}}
           \log{s\over t} \nonumber \\
     & + &{{(1-2c_W^2)^3}\over{c_W^2s_W^2}}(\log {u\over t})(\log{s\over{M_Z^2}})
        -2{{1-2c_W^2}\over{s_W^2}}~{s\over u}(\log{s\over t})(\log{s\over u})
            \nonumber \\
     & - &\Big[ {{1-16c_W^2+20c_W^4}\over{4c_W^2s_W^2}}~{s\over u}
            +{{1-2c_W^2}\over{4c_W^2s_W^2}}s{{s+3t-6c_W^2s}
\over{t^2+u^2}}\Big]
              (\log{s\over t})^2 \nonumber \\
     & - &( {1\over{4c_W^2s_W^2}}~{s\over t}
            +{{1-2c_W^2}\over{2s_W^2}}~{{3st}\over{t^2+u^2}} )(\log{s\over u})^2
            \nonumber \\
     & - &4s_W^2(\log{u\over t})^2 -16s_W^2 Sp(-{u\over t})
         -{{1-2c_W^2}\over{4c_W^2s_W^2}} s{{3s+t+6c_W^2(s+3t)}\over{t^2+u^2}}
           \log{s\over t} \nonumber \\
     & - &{{(1-2c_W^2)(1-6c_W^2)}\over{4c_W^2s_W^2}}~{{su}\over{t^2+u^2}}
         + {3\over 2}~ {{m_t^2}\over{s_W^2M_W^2}}
\log{{m_t^2}\over s} \Biggr]
         +0.030, 
\end{eqnarray}

\begin{eqnarray}
     S_Q^{(+)dom} & = &{{\alpha}\over{4\pi}}
      \Biggl[ -{{5s_W^4+3c_W^4}\over{4c_W^2s_W^2}} (\log{s\over {M_W^2}})^2
          + {{65s_W^2+18c_W^4}\over{6c_W^2s_W^2}}\log{s\over {M_W^2}}
\\
     & - & {{(1-2s_W^2)^2+4s_W^4}\over{2c_W^2s_W^2}}\log c_W^2~\log{s\over{M_W^2}} \nonumber \\
     & + &2{{1-2c_W^2}\over{c_W^2}}\log{u\over t} \log{s\over{M_Z^2}}
         +{s\over{2c_W^2 u}}(\log{s\over t})^2  \nonumber \\
     & - & {s\over{2c_W^2 t}}(\log{s\over u})^2 -2(\log{u\over t})^2
         -8Sp(-{u\over t}) + {{3m_t^2}\over{2s_W^2 M_W^2}}\log{{m_t^2}\over s}
         \Biggr] + 0.045.\nonumber
\end{eqnarray}
\normalsize

\begin{figure}[p]
\begin{center}
\epsfig{file=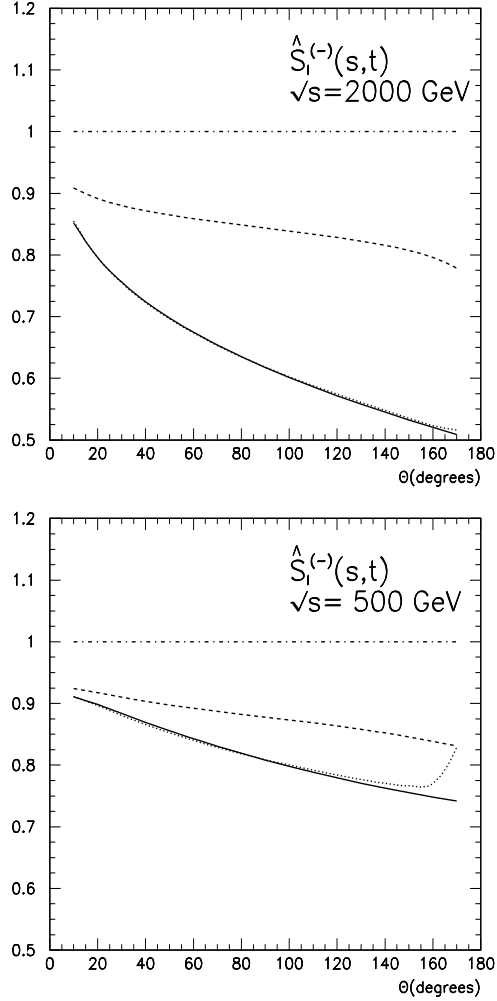,width=8cm,height=14.5cm}
\end{center}
\caption{The Born-form invariant amplitude $\hat S^{(-)}_l (s,t)$ as a
function of the W-production angle, $\theta$, for $\sqrt s = 2000 GeV$ and
$\sqrt s = 500 GeV$ in (i) the full one-loop evaluation including 
soft-photon bremsstrahlung (solid line), (ii) the fermion-loop approximation
including soft-photon bremsstrahlung (dashed line), (iii) the high-energy
approximation based on (14) to (16) (dotted line), (iv) the Born approximation
(dash-dotted line).}
\end{figure}

\begin{figure}[p]
\begin{center}
\epsfig{file=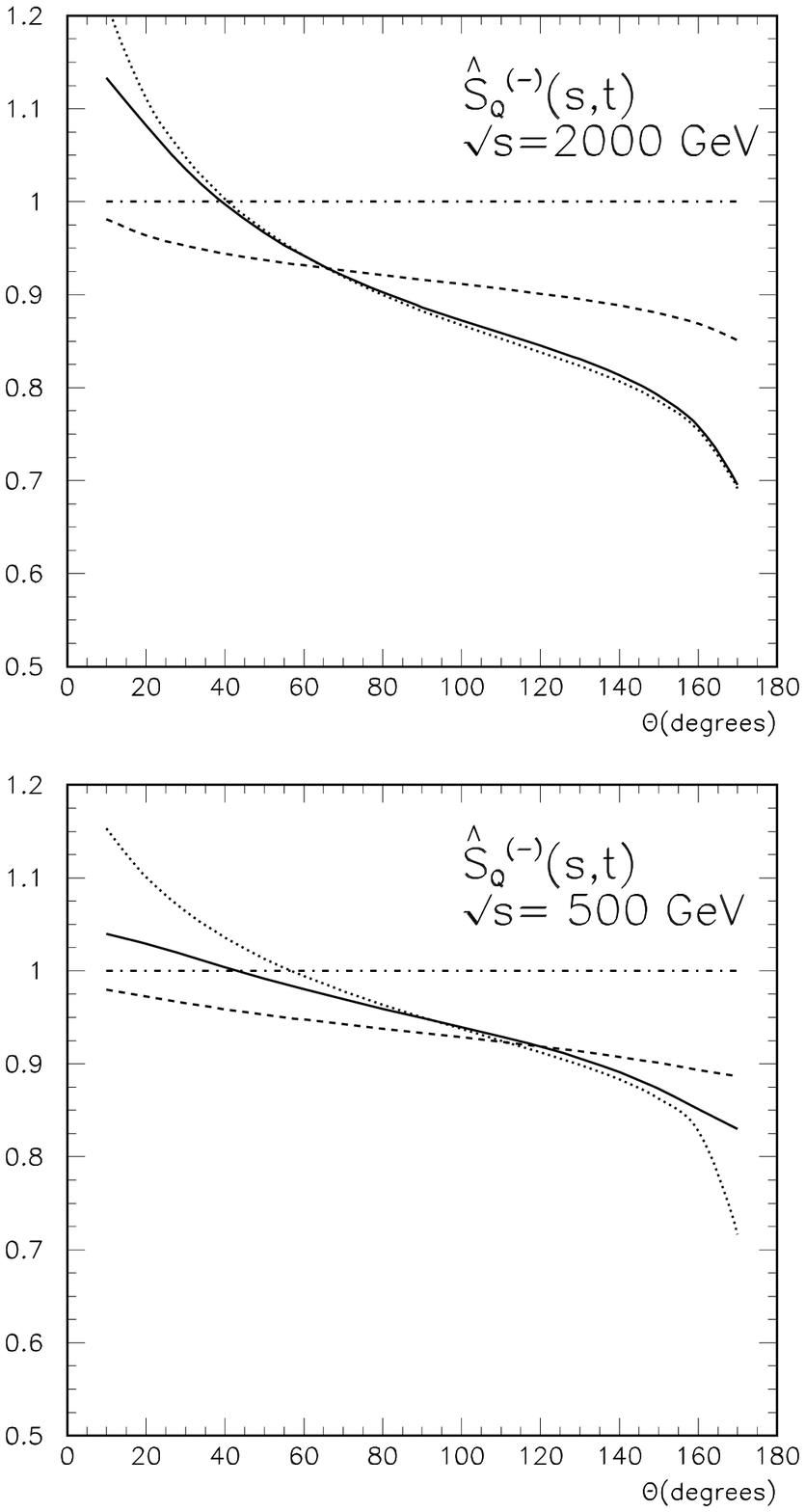,width=10cm,height=16.5cm}
\end{center}
\caption{Same as Fig. 1, but for  $\hat S^{(-)}_Q (s,t)$}
\end{figure}

\begin{figure}[p]
\begin{center}
\epsfig{file=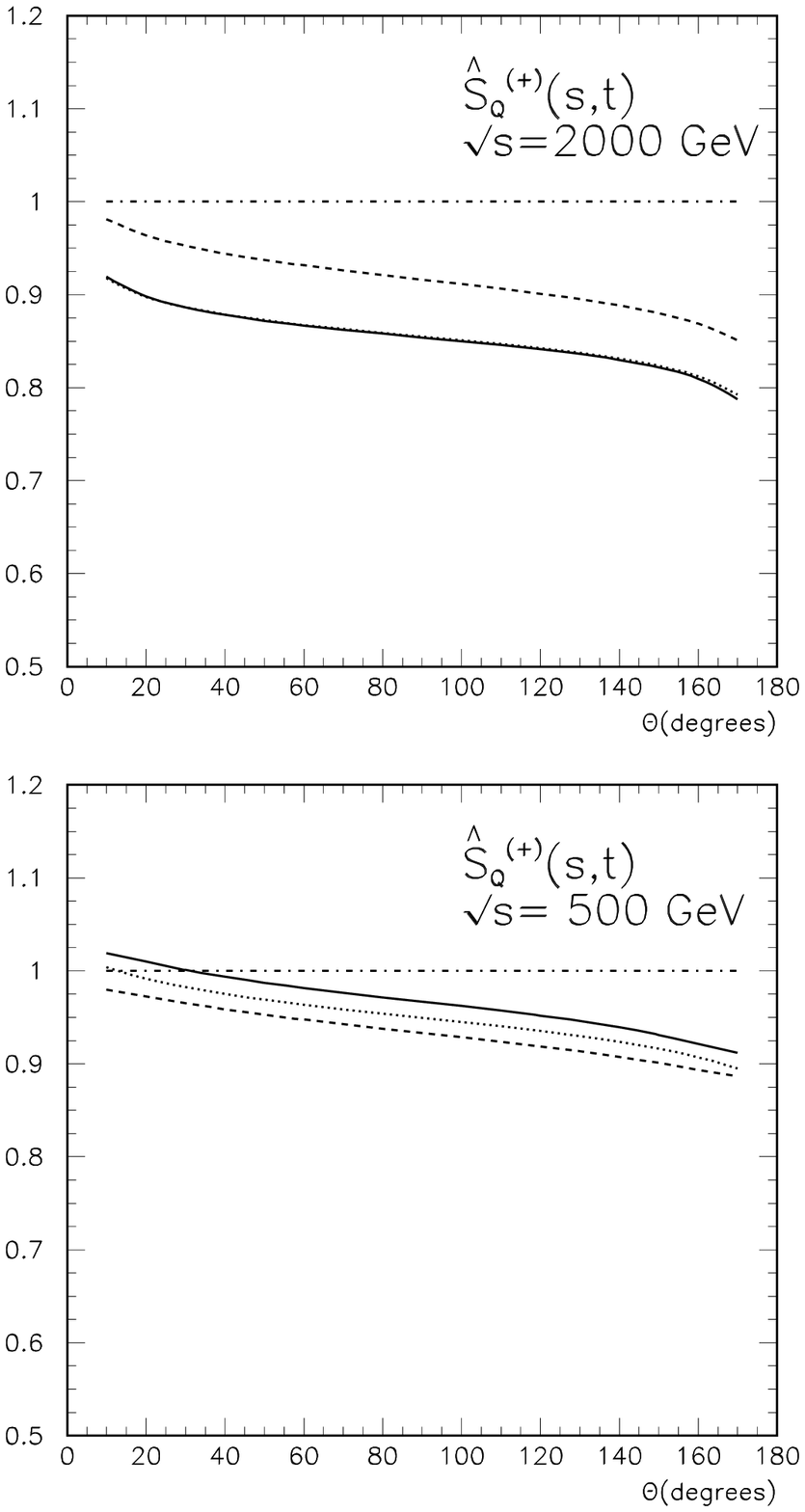,width=10cm,height=16.5cm}
\end{center}
\caption{Same as Fig. 1, but for  $\hat S^{(+)}_Q (s,t)$}
\end{figure}

In figs. 1 to 3, \cite{Ku-Schi},
for $\sqrt s = 500$ GeV and 2000 GeV, I show the invariant
amplitudes $\hat S^{(-)}_I$ and $\hat S^{(\pm)}_Q$ entering the HEBFA.
The soft-photon cut-off is chosen as $\Delta E = 0.025 \sqrt s$. We note
that over much of the angular range of the production angle the HEBFA yields
a very good approximation of the full one-loop results. The quality of the
approximation (obviously) improves with increasing energy. In figs. 1 to 3,
we also indicate the results obtained for $\hat S^{(-)}_I$ and $S^{(\pm)}_Q$
if only fermion loops and soft-photon radiation is taken into account. The
remarkably large difference between the results with only fermion loops
and the full corrections is an important genuine effect of electroweak
loop corrections. Its large magnitude is due to the squared (non-Abelian
Sudakov) logs appearing in the expressions (14) to (16).

We turn to the accuracy of the total cross section, when evaluated in
HEBFA. In Table 2, we present the accuracy $\Delta (\%)$ defined by
\beq
\Delta (\%) = {{d \sigma_{appr.} - d \sigma_{full~ one-loop}} \over
{d \sigma_{Born}}}.
\eeq
Table 2 first of all shows the accuracy of the Born-form approximation, i.e. 
dropping all terms beyond the Born form in (12), but evaluating $\hat S_I$
and $\hat S_Q^{(\pm)}$ at one loop exactly. Secondly, Table 2 shows the result
of using the HEBFA for $\hat S^{(-)}_I$ and $\hat S^{(\pm)}_Q$.
We conclude that the accuracy of the HEBFA is
excellent, except for the case of mixed polarizations of the W bosons,
which is strongly suppressed in magnitude, however. For a detailed
discussion of the results for angular distributions, we refer to
ref. \cite{Ku-Schi}.

\begin{table}[h]
\scriptsize
\begin{center}
\begin{tabular}{|r||l|r||l|r||l|l|}\hline
   angle & \multicolumn{2}{c||}{High-energy-Born-form} 
         & \multicolumn{2}{c||}{Born-form}
         & \multicolumn{1}{c|}{full~one-loop}
         & \multicolumn{1}{c|}{Born} \\
         & \multicolumn{2}{c||}{approximation}
         & \multicolumn{2}{c||}{approximation}
         &    &   \\ \hline
         & \multicolumn{1}{c|}{$\sigma(pb)$}
         & \multicolumn{1}{c||}{$\Delta(\%)$}
         & \multicolumn{1}{c|}{$\sigma(pb)$}
         & \multicolumn{1}{c||}{$\Delta(\%)$}
         & \multicolumn{1}{c|}{$\sigma(pb)$}
         & \multicolumn{1}{c|}{$\sigma(pb)$}
            \\  \hline \hline
     & \multicolumn {6}{c|}{$\sqrt s = 2000$ GeV } \\ \hline
  ``unpol.''
     & 1.461$\times 10^{-1}$ & +0.14 & 1.461$\times 10^{-1}$  &  +0.16 
     & 1.457$\times 10^{-1}$         & 2.758$\times 10^{-1}$ \\ \hline 
  transv.
     & 1.422$\times 10^{-1}$ & +0.19 & 1.423$\times 10^{-1}$  &  +0.19 
     & 1.417$\times 10^{-1}$         & 2.683$\times 10^{-1}$ \\ \hline 
  longit.
     & 3.526$\times 10^{-3}$ &$-$0.10& 3.533$\times 10^{-3}$  &  0.00 
     & 3.533$\times 10^{-3}$         & 6.788$\times 10^{-3}$ \\ \hline 
  mixed
     & 2.912$\times 10^{-4}$&$-$14.79& 2.909$\times 10^{-4}$ &$-$14.83 
     & 3.833$\times 10^{-4}$         & 6.229$\times 10^{-4}$ \\ \hline 
       \hline\hline
     & \multicolumn {6}{c|}{$\sqrt s = 500$ GeV } \\ \hline
  ``unpol.''
     & 3.448 &$-$0.42& 3.462 &$-$0.11 & 3.467  & 4.545 \\ \hline 
  transv.
     & 3.260 &$-$0.34& 3.274 &$-$0.01 & 3.274  & 4.294 \\ \hline 
  longit.
     & 8.287$\times 10^{-2}$ &$-$0.33& 8.323$\times 10^{-2}$  &  0.00 
     & 8.323$\times 10^{-2}$         & 1.091$\times 10^{-1}$ \\ \hline 
  mixed
     & 1.033$\times 10^{-1}$ &$-$3.19& 1.034$\times 10^{-1}$  &$-$3.13 
     & 1.078$\times 10^{-1}$         & 1.419$\times 10^{-1}$ \\ \hline 
       \hline
\end{tabular} 
\end{center}
\caption{
The total  cross section  for  $W^+W^-$ pair production 
(obtained by integration over the angular range of the production angle of
$10^\circ \leq \vartheta \leq 170^\circ$)
at $\sqrt s= 2000$ GeV and $\sqrt s= 500$ GeV. The different rows show the 
results when summing over the $W^+W^-$ spins (``unpol.'') and the results for
the various cases of polarization of the produced $W^+$ and $W^-$. 
The first column shows the result of
the HEBFA based on (13) to (16).  
The second column gives the result of the
Born-form approximation obtained  by
evaluating $\hat S^{(-)}_I$ and $\hat S^{(\pm)}_Q$ at 
one-loop level exactly. The
third column shows the full one-loop result and the Born approximation. 
}
\end{table}

\section{HEBFA for \boldmath $e^+ e^- \to W^+ W^- \to 4 {\rm fermions} 
(+ \gamma)$\unboldmath
at one loop.}
In a very recent paper \cite{Kuri}, the HEBFA was supplemented by including
the decay of the W bosons and hard-photon radiation. Specifically, we looked
at the decay channel
\beq
e^+ e^- \to W^+ (u \bar d) W^- (\bar s c) (+ \gamma),
\eeq
as well as the semileptonic channel
\beq
e^+ e^- \to W^+ (u \bar d) W^- (e \bar \nu) (+ \gamma).
\eeq

A two-step procedure was employed in ref. \cite{Kuri}. In a first step,
we showed that the background of four-fermion production not proceeding
via two W bosons can be suppressed by an appropriate cut,
\beq
\left| \sqrt{k^2_\pm} - M_W \right| \leq 5 \Gamma_W,
\eeq
on the invariant mass $\sqrt{k^2_\pm}$ of the produced fermion pairs. In a
tree-level calculation, using GRACE \cite{GRACE}, we compared the production
of four fermions via intermediate W bosons with the production process
\beq
e^+ e^- \to u \bar d \bar c s
\eeq
based on the full set of all contributing diagrams. While in general the
introduction of Breit-Wigner denominators in the process (21) leads to
problems of gauge invariance (e.g. \cite{Been}), the ``fixed-width scheme''
employed for the background estimate finds some justification in a 
``complex-mass scheme'' \cite{Roth} and should be sufficiently reliable.
The results in Table 3, \cite{Kuri}, in particular a 
comparison of lines 5 and 8, show
that the cut (20) on the invariant masses of the fermion pairs removes
the non-doubly-resonant background to the level of less than 0.3 \%. The
suppression of the background in the semileptonic channel is only slightly
larger. The results on $\Delta$, corresponding to the last line in table 3, are
given by 0.9 \%, 0.4 \% and 0.4 \%, respectively.

\begin{table}[h]
\scriptsize
\begin{tabular}{|r|l|c|c|c|}\hline
Line & $\sqrt s$ & 500 GeV & 1 TeV & 2 TeV\\ \hline\hline
 1&$\sigma (e^+ e^- \to W^+W^-)$ & 7.458 & 2.785 & $9.421 \times 10^{-1}$\\ \hline\hline
\multicolumn{5}{|c|}{Zero width approximation}\\ \hline
2&$\sigma \times BR (W^+ \to u \overline{d}) \times BR (W^- \to \overline{c}s)$
&$8.289 \times 10^{-1}$ & $3.094\times 10^{-1}$ &
 $1.047 \times 10^{-1}$ \\ \hline\hline
\multicolumn{5}{|c|}{Breit-Wigner, full four-fermion phase space} \\ \hline
3&$\sigma (e^+ e^- \to W^+ (\to u \overline{d}) W^- (\to \overline{c} s))$
&$8.291 \times 10^{-1}$ & $3.097 \times 10^{-1}$ &
 $1.046 \times 10^{-1}$ \\
4&$\sigma (e^+ e^- \to u \overline{d} \overline{c} s)$ &
$8.466 \times 10^{-1}$ & $3.248 \times 10^{-1}$ &
 $1.124 \times 10^{-1}$ \\
5&Difference $\Delta$ in \% & 2.1 \% & 4.9\% & 7.5\% \\ \hline\hline
 \multicolumn{5}{|c|}{Breit-Wigner, restricted phase space, 
$\vert \sqrt{k^2_\pm} - M_W \vert \lsim 5 \Gamma_W$} \\ \hline
6&$\sigma (e^+ e^- \to W^+ (\to u \overline{d}) W^- (\to \overline{c} s))$
&$7.264 \times 10^{-1}$ & $2.713 \times 10^{-1}$ &
$9.16 \times 10^{-2}$ \\
7&$\sigma (e^+ e^- \to u \overline{d} \overline{c} s)$ & 
$7.275 \times 10^{-1}$ & $2.717 \times 10^{-1}$  & 
$9.19 \times 10^{-2}$ \\
8&Difference $\Delta$ in \% & 0.1 \% & 0.1 \% & 0.3 \% \\ \hline
\end{tabular}
\caption{Tree-level results in [pb] for $W^+W^-$-mediated four-fermion
production (specifically for the $u \overline{d} \overline{c} s$ final
state) compared with four-fermion production including (non-doubly-resonant)
background for different phase-space cuts.}
\end{table}

\begin{table}[tbp]
\scriptsize
\begin{tabular}{|r|c|c|c|c|r|}\hline
              & $e^+e^-\to W^+W^-$ & $e^+e^-\to W^+(u\bar d)W^-(\bar c s)$
              & \multicolumn{3}{c|}
                {$e^+e^-\to W^+(u\bar d)W^-(\bar c s)+\gamma$} \\ \cline{2-6}
              & Born &  Born & \multicolumn{3} {c|}{one-loop}  \\ \cline{4-6}
  $\cos\theta$   &    &   & HEBFA  &  exact & $\Delta (\%)$ \\ \hline
  0.95& $5.981\times 10^0  $  & $5.827\times 10^{-1}$ &
        $2.900\times 10^{-1}$ & $2.878\times 10^{-1}$ & 0.76 \\ 
  0.9 & $2.785\times 10^0  $  & $2.713\times 10^{-1}$ &
        $1.211\times 10^{-1}$ & $1.208\times 10^{-1}$ & 0.24 \\ 
  0.8 & $1.207\times 10^0  $  & $1,176\times 10^{-1}$ &
        $4.557\times 10^{-2}$ & $4.557\times 10^{-2}$ & 0.00\\
  0.7 & $7.003\times 10^{-1}$ & $6,826\times 10^{-2}$ &
        $2.383\times 10^{-2}$ & $2.385\times 10^{-2}$ & -0.05\\
  0.6 & $4.597\times 10^{-1}$ & $4.483\times 10^{-2}$ &
        $1.437\times 10^{-2}$ & $1.438\times 10^{-2}$ & -0.02\\
  0.5 & $3.246\times 10^{-1}$ & $3.165\times 10^{-2}$ &
        $9.429\times 10^{-3}$ & $9.435\times 10^{-2}$ & -0.06 \\
  0.4 & $2.414\times 10^{-1}$ & $2.352\times 10^{-2}$ &
        $6.570\times 10^{-3}$ & $6.576\times 10^{-3}$ & -0.10 \\
  0.3 & $1.869\times 10^{-1}$ & $1.821\times 10^{-2}$ &
        $4.798\times 10^{-3}$ & $4.808\times 10^{-3}$ & -0.20 \\
  0.2 & $1.497\times 10^{-1}$ & $1.458\times 10^{-2}$ &
        $3.645\times 10^{-3}$ & $3.651\times 10^{-3}$ & -0.17\\
  0.1 & $1.234\times 10^{-1}$ & $1.201\times 10^{-2}$ &
        $2.855\times 10^{-3}$ & $2.861\times 10^{-3}$ & -0.22 \\
  0.0 & $1.041\times 10^{-1}$ & $1.013\times 10^{-2}$ &
        $2.292\times 10^{-3}$ & $2.297\times 10^{-3}$ & -0.23 \\
 -0.1 & $8.941\times 10^{-2}$ & $8.695\times 10^{-3}$ &
        $1.872\times 10^{-3}$ & $1.876\times 10^{-3}$ & -0.22 \\
 -0.2 & $7.766\times 10^{-2}$ & $7.551\times 10^{-3}$ &
        $1.542\times 10^{-3}$ & $1.544\times 10^{-3}$ & -0.16 \\
 -0.3 & $6.773\times 10^{-2}$ & $6.586\times 10^{-3}$ &
        $1.268\times 10^{-3}$ & $1.269\times 10^{-3}$ & -0.05 \\
 -0.4 & $5.883\times 10^{-2}$ & $5.721\times 10^{-3}$ &
        $1.031\times 10^{-3}$ & $1.030\times 10^{-3}$ &  0.06 \\
 -0.5 & $5.036\times 10^{-2}$ & $4.897\times 10^{-3}$ &
        $8.174\times 10^{-4}$ & $8.148\times 10^{-4}$ &  0.31 \\
 -0.6 & $4.188\times 10^{-2}$ & $4.073\times 10^{-3}$ &
        $6.202\times 10^{-4}$ & $6.155\times 10^{-4}$ &  0.77 \\
 -0.7 & $3.305\times 10^{-2}$ & $3.214\times 10^{-3}$ &
        $4.364\times 10^{-4}$ & $4.291\times 10^{-4}$ &  1.70 \\
 -0.8 & $2.360\times 10^{-2}$ & $2.295\times 10^{-3}$ &
        $2.680\times 10^{-4}$ & $2.573\times 10^{-4}$ &  4.14 \\
 -0.9 & $1.333\times 10^{-2}$ & $1.296\times 10^{-3}$ &
        $1.215\times 10^{-4}$ & $1.074\times 10^{-4}$ & 13.19 \\ \hline
\end{tabular}
\caption{The angular distribution  of W-pair production 
at the  energy $\sqrt s =2E_{beam} = 1$ TeV in units of $pb$.  
The first column shows  the Born cross section for $e^+e^-\to W^+W^-$. 
The second column shows the results of treating W production and decay
in Born approximation and integrating the Breit-Wigner distribution
over the restricted interval (20).
The third and the fourth column are obtained by using the one-loop
amplitudes for production and decay, again, integrating the Breit-Wigner
distribution over the restricted interval (20).
A soft-photon cut $\Delta E/E= 0.01$ is used for the one-loop
results.  The HEBFA is used for the third column and the full one-loop
amplitudes are  used for the fourth column.  The last column gives the
results for the relative deviation, $\Delta$, from (22).
}
\end{table}
\begin{table}[tbp]
\scriptsize
\begin{tabular}{|r|c|c|c|c|c|}\hline
              & $e^+e^-\to W^+W^-$ & $e^+e^-\to W^+(u\bar d)W^-(\bar c s)$
              & \multicolumn{3}{c|}
                {$e^+e^-\to W^+(u\bar d)W^-(\bar c s)+\gamma$} \\ \cline{2-6}
               & Born    &  Born  &\multicolumn{3}{c|}{one-loop}\\ \cline{4-6}
 $E_{\rm beam}$&         &        & HEBFA  &  exact  & $\Delta(\%)$ 
                                                     \\ \hline
  200 & 8.698  & 0.8467  & 0.8718 & 0.8794 &-0.9  \\ 
  300 & 5.091  & 0.4958  & 0.5275 & 0.5262 & 0.2  \\ 
  400 & 3.384  & 0.3295  & 0.3575 & 0.3533 & 1.2  \\ 
  500 & 2.433  & 0.2370  & 0.2602 & 0.2556 & 1.8  \\ 
  600 & 1.844  & 0.1795  & 0.1996 & 0.1951 & 2.3  \\ 
  700 & 1.452  & 0.1413  & 0.1584 & 0.1543 & 2.7  \\ 
  800 & 1.177  & 0.1145  & 0.1292 & 0.1259 & 2.6  \\ 
  900 & 0.9750 & 0.09485 & 0.1080 & 0.1044 & 3.4  \\ 
 1000 & 0.8228 & 0.08010 & 0.0915 & 0.0881 & 3.9  \\ \hline
\end{tabular}
\caption{The energy dependence of the $(u\bar d)(\bar c s)$- 
production cross section in DPA. The second column is the Born cross section,
while the third column gives  the  one-loop cross section 
including hard-photon radiation. The deviation, $\Delta$, defined in
analogy to (22), quantifies the discrepancy between the HEBFA and the
full one-loop results. 
}
\end{table}

\vspace{0.5 cm}
\begin{table}[tbp]
\scriptsize
\begin{tabular}{|r|c|c|c|c|c|}\hline
              & $e^+e^-\to W^+W^-$ & $e^+e^-\to W^+(u\bar d)W^-(\bar c s)$
              & \multicolumn{3}{c|}
                {$e^+e^-\to W^+(u\bar d)W^-(\bar c s)+\gamma$} \\ \cline{2-6}
               & Born    &  Born   &\multicolumn{3}{c|}{one-loop}\\ \cline{4-6}
 $E_{\rm beam}$&         &         & HEBFA   &  exact  & $\Delta(\%)$
                                                        \\ \hline
  200 & 6.724  & 0.6561  & 0.6746  & 0.6794  &-0.71  \\
  300 & 3.042  & 0.2964  & 0.3109  & 0.3109  & 0.00  \\
  400 & 1.695  & 0.1654  & 0.1725  & 0.1721  & 0.23  \\
  500 & 1.077  & 0.1051  & 0.1085  & 0.1082  & 0.28  \\
  600 & 0.7440 & 0.07262 & 0.07405 & 0.07383 & 0.30  \\
  700 & 0.5449 & 0.05318 & 0.05349 & 0.05334 & 0.28  \\
  800 & 0.4162 & 0.04063 & 0.04027 & 0.04015 & 0.30  \\
  900 & 0.3284 & 0.03205 & 0.03136 & 0.03127 & 0.29  \\
 1000 & 0.2657 & 0.02593 & 0.02505 & 0.02498 & 0.28  \\ \hline
\end{tabular}
\par
\medskip
\caption{As Table 5, but with 
a restriction on the $W^+W^-$ production angle that is given by 
$10^\circ< \theta <170^\circ$. 
}
\end{table}

We turn to the second step, the calculation of the cross sections for
reactions (18), (19) at one loop at collider energies. Extensive calculations
have demonstrated \cite{Fadin,De-Di-Ro} that non-factorizable corrections
to four-fermion production are small at high energies, as one might expect,
and moreover, they vanish upon integration over the invariant masses of the 
fermion pairs. Accordingly, it is justified to employ one-loop 
W-pair-production and -decay amplitudes, when evaluating reactions (18) and
(19) at one-loop level. Moreover, non-doubly-resonant contributions are 
suppressed by imposing the cut (20).

In detail, the numerical results \cite{Kuri} to be presented are based on
\begin{itemize}
\item[i)] one-loop on-shell $W^+W^-$ production and decay amplitudes, based
on the full one-loop results from \cite{Analytic} as well as the HEBFA from
\cite{Ku-Schi},
\item[ii)] fixed-width Breit-Wigner denominators and the phase-space
cut (20), i.e. a double-pole approximation with respect to four-fermion
production,
\item[iii)] inclusion of hard-photon emission generated by GRACE \cite{GRACE}
and the Monte Carlo routine BASES \cite{Kawa}, 
\item[iv)] independence of the soft-photon cut-off $\Delta E$ for
$1 GeV < \Delta E < 10 GeV$.
\end{itemize}

With canonical values for the input parameters, $M_Z = 91.187 GeV,~M_W =
80.22 GeV,~M_H = 200 GeV$, the results in Tables 4 to 6 were obtained.

Table 4 demonstrates that indeed at $\sqrt s = 1 TeV$, the deviation
\beq
\Delta = {{{{d \sigma} \over {d \cos \vartheta}} (HEBFA) - 
{{d \sigma} \over {d \cos \vartheta}} (exact)} \over {{{d \sigma} \over 
{d \cos \vartheta}} (exact)}} < 0.5 \% \nonumber
\eeq
is less than 0.5 \%, except for very forward and very backward production
angles $\vartheta$. Finally, Table 5 and Table 6 show the energy 
dependence of the
total cross section. Upon applying the angular cut, $10^0 < \vartheta <
170^0$, the accuracy of the total cross section becomes better than
0.3 \% for c.m.s. energies above 500 GeV.

Finally, comparing the results of taking into account only fermion
loops and photon radiation with the results from the full one-loop
calculation, one finds \cite{Kuri}
differences that reach approximately 20 \% at
2 TeV c.m.s. energy. Accuracies of future experiments of this order of
magnitude will accordingly be able to ``see'' the non-Abelian loop corrections
displayed in (14) to (16).

\section{Conclusions}
The main points of this review may be summarized as follows:
\begin{itemize}
\item[i)] Concerning the LEP2 energy range, the simple procedure of introducing
the SU(2) gauge coupling $g_{W^\pm} (M_W^2)$ at the high-energy scale, 
approximated by the W-mass-shell condition
$s \simeq M_W^2$, and the electromagnetic coupling $\alpha
(M^2_Z)$, allows one to incorporate most of the electroweak virtual radiative
corrections to $e^+e^- \to W^+W^-$ in a simple Born formula.
\item[ii)] The detailed numerical results obtained at tree-level at high
energies show that a cut of about five times the $W$ width on fermion-pair
masses enhances production via W-pairs, reducing non-resonant background
to below 0.2 \% for $e^+e^- \to W^+ (u \bar d) W^- (\bar c s)$ and below
0.4 \% for $e^+e^- \to W^+ (u \bar d) W^- (e^- \bar \nu)$. It is accordingly
sufficient to concentrate on $e^+e^- \to W^+ W^- \to 4 {\rm fermions}$
(i.e. the double-pole approximation) and ignore background contributions,
even more so, as in four-fermion production the main interest lies in the
test of the non-Abelian gauge-boson interactions of the electroweak
theory.
\item[iii)] The HEBFA is excellent for $\sqrt s \gsim 400 GeV$, provided 
very-forward and very-backward production is excluded. It is 
conceptually simple, its analytic expressions fit on two pages, and it
is practically important due to a significant reduction in computer time
in comparison with the full one-loop calculation.
\item[iv)] Accuracies of future experiments of the order of magnitude of
10 \% in the total cross section at TeV energies
allow one to isolate bosonic loop
corrections.
\end{itemize}


\end{document}